\documentstyle[12pt]{article}
\title{\begin{flushright}
\normalsize{hep-th/9311003}
\end{flushright}
\vspace{.5cm}
Theory of Superselection Sectors for Generalized Ising models}
\author{J.R.Reyes Mart\'{\i}nez\thanks{risc4@physik.fu-berlin.de} \\
 Institut f\"{u}r Theoretische Physik, FU-Berlin \\
  Arnimallee 14, D-14195 Berlin }
\date{October 1993}
\begin{document}
\maketitle
\begin{abstract}

     	We apply the theory of superselection sectors in the same way as
done by G.Mack and V.Schomerus for the Ising model to generalizations
of this model described by J.Fr\"{o}hlich and T.Kerler.

\end{abstract}
\newpage

\newcommand{\be}{\begin{equation}}
\newcommand{\ee}{\end{equation}}
\newcommand{\bea}{\begin{eqnarray}}
\newcommand{\eea}{\end{eqnarray}}
\newcommand{\nn}{\nonumber}
\newcommand{\bean}{\begin{eqnarray*}}
\newcommand{\eean}{\end{eqnarray*}}
\newcommand{\ba}{\begin{array}}
\newcommand{\ea}{\end{array}}


\newcommand{\ai}{{\alpha}_{i}}
\newcommand{\ep}{\varepsilon}
\newcommand{\ag}{{\alpha}_{g}}
\newcommand{\fum}{\frac{1}{2}}
\newcommand{\ra}{\rangle}
\newcommand{\fr}{\frac}
\newcommand{\pns}{{\pi}_{NS}}
\newcommand{\pr}{{\pi}_{R}}
\newcommand{\Mns}{{\sf Maj}_{NS}}
\newcommand{\Mr}{{\sf Maj}_{R}}
\newcommand{\ro}{{\rho}_{_{1/2}}}
\newcommand{\ii}{\underline i}
\newcommand{\eg}{E_{g}}
\newcommand{\med}{\protect{\textstyle{\frac{1}{2}}}}

\section{Introduction}

	In the framework of Algebraic Field Theory the endomorphisms
of the observable algebra play a fundamental role, they describe the
different sectors (irreducible representations of the observables).
The fusion rules satisfied by these endomorphisms are expected to
correspond to the tensor product of the representations of the unknown
symmetry object.  In four space-time dimensions we have permutation
group statistics and the symmetry is realized by compact groups, in
2-dim we have braid group statistics and the corresponding symmetry is
not in general known (3-dim is a intermediary case where the two
possibilities can generically occur depending on the localization of
the representations allowed).  Conformal field theories in 2-dim are
an exellent testing ground for unreveling this new symmetry structure,
there are plenty of models where we know the sectors and their fusion
rules completely.  Unfortunately knowing the composition of the
endomorphisms abstractly, i.e. knowing their fusion rules only, has
not been very helpful in the quest for this new concept (another point
is that the fusion rules do not determine a specific model uniquely).
It seems almost impossible to explicitly construct localized
endomorphisms (localization of the representations is important to see
the braid group statistics see for example \cite{frs1}), but all we
need in principle are endomorphisms unitary equivalent to local ones.
This was done in \cite{ms} and
\cite{fgv}, where the morphisms constructed,
although {\em not} localized , enabled a very complete analysis of the
models in question.

Coming now to
the proper subject of our paper, we  present a class of models
where we can construct the endomorphisms in the same spirit as done
for the Ising model by Mack and Schomerus.
 The models that we analyze
were classified in \cite{fk}. They have the fusion rules:

\bea
      \ro\circ\ro&=&\!\!\sum_{\ag\in stab(\ro)}\ag \label{fu1}\\
			\ag\circ\ro&=&\ro \label{fu2}
\eea

	Where the second equation expresses the fact that
\mbox{$\ag\in stab(\ro)$}. The only possible groups, $stab(\ro)$,
that can occur ( so that the intertwiners between sectors and products
of sectors satisfy the braid and fusion identities )
 have $M$ inequivalent self-conjugating
automorphirms, i.e., $stab(\ro)\cong({\bf Z}_{2})^{M}$ then the above
sum runs over $2^M$ terms and the statistical dimension of $\ro$ is
$d_{\ro}\!= 2^{\fr{M}{2}}$.  To simplify the discussion the
endomorphism $\ro$ was chosen to be selfconjugate. To realize these
models we start with $M$ Majorana fields.


\section{The Lie algebra of observables its \protect{\\}
          representations and morphims}


\subsection{Representations}

	As promised we define:

	\[{\sf Maj} = \sum_{i=1}^{M}{\sf Maj}_{{NS}_{(i)}}
	\oplus{\sf Maj}_{{R}_{(i)}}\]

\bean
  {\sl b}_{a,i} \quad{\rm for} & a \in{\bf Z} + \fum & \rightarrow
  {\sf Maj}_{{NS}_{(i)}} \\
  {\sl b}_{a,i}  \quad{\rm for} & a \in{\bf Z}  & \rightarrow
  {\sf Maj}_{{R}_{(i)}}
\eean

 	The anticommutation relations are:

\bea
 \{\sl b_{a,i},\sl b_{c,j}\} = \fum[1+(-1)^{2a} Y]\delta_{i,j}
 \delta_{a,-c};\qquad b^{\ast}_{a,i}=b_{-a,i}
\eea

	Where the $Y$ commute with all fermion operators and among
themselves,i.e., it belong to the center of the algebra $\sf Maj$.
Note also that for every $i$, $Y=4b_{0,i}b_{0,i}-1$ and
$Yb_{a,i}=(-1)^{2a}b_{a,i}$.

 Our observable algebra must contain the
Virasoro algebra as a sub-algebra whose generators are given in terms of
the Majorana operators as follows:

\bean
{\sl L_{n}=\sum_{i=1}^{M}(\sum_{c>\frac{n}{2}}(c - \frac{n}{2})b_{n-c,i}
  b_{c,i} + \frac{1}{8}\delta_{n,0}b_{0,i}^{2})}\quad
	;\mbox{  for  $n \geq 0$}
\eean

	The action of $\ro$ (see below)
on ${\it Vir}$ will take us out of this
algebra therefore we enlarge our observable algebra ${\cal A}$
so that ${\ro}$ is a properly defined endomorphism
(because $\ro$ turns out to be blind on 'colors'
${\cal A}\supset{\it Vir}$ is now
generated by similar bilinears
as in \cite{ms}), namely

\bea
	 L_{n} , N_{n} \quad    (n \in \bf Z)  \nn
\eea
\bea
 J_{ab,i} = -J_{ba,i}\hspace{.5cm} (a,b \in \fum{\bf Z},a-b\in{\bf Z}
	\setminus \!\{0\})\qquad \forall i\in\{1,2,...,M\}
\eea

	Under hermitian conjugation we have:

\bea
    L_{n}^{\ast} =L_{-n};\; N_{n}^{\ast} = N_{-n};\;
	J_{ab,i}^{\ast} =J_{-b-a,i}
\eea

	The generators $\sl N_{n}$ and $\sl J_{ab,i}$ are given by:

\bea
   N_{n} = \sum_{c > n}\sum_{i=1}^{M}b_{n-c,i}b_{c,i}\quad for \; n \geq0
\eea

\bea
   J_{ab,i}=b_{a,i}b_{c,i}- \frac{1}{4}[1 +(-1)^{2a}Y]\delta_{a,-c}
\eea

	These generators close under commutation relations because we
are only taking summations or terms with diagonal 'i' components and
we can use Mack and Schomerus calculations (prop. 2.8) taking now sums
over 'i' when required. The global observable algebra $\bar {\cal A}$
is now the envelopping algebra of $\cal A$. The central charge $c$ is
equal to $\frac{M}{2}$.

	We now define two  irreducible $\ast$-representations
of $\sf Maj$ denoted $\pns$ and $\pr$ by:

\bean
	 \pns(b_{a,i})=0&{\rm for\; a} \in{\bf Z}\quad\pns(Y)\!&=-1\\
	\pr(b_{a,i})=0&{\rm for\; a} \in {\bf Z}+\fum\quad\pr(Y)&=1
\eean

	They obey:

\[ {\sf Maj}/Ker(\pns)\cong  \Mns  , {\sf Maj}/Ker(\pr)\cong  \Mr \]

	The corresponding modules $\cal H_{NS}$ and $\cal H_{R}$ are highest
weight modules:

\bean
	b_{a,i}|NS\ra=0&{\rm for\;} b_{a,i}\in \Mns& \; ,a>0 \\
	b_{a,i}|R\ra=0&{\rm for\;} b_{a,i}\in \Mr& \; ,a>0
\eean

	$\cal H_{NS}$ and $\cal H_{R}$ are spanned by the vectors:
\bea
{}&|(\vec i \vec a)_{m}{\ra}_{NS}& \equiv |i_{1}a_{1},i_{2}a_{2},\ldots,
i_{m}a_{m}\ra_{NS}:=b_{-a_{m},i_{m}}\ldots b_{-a_{2},i_{2}}b_{-a_{1},i_{1}}
{|NS\ra}_{NS}  \label{vec1} \\
{}&|(\vec i \vec a)_{m}{\ra}_{R}&\equiv |i_{1}a_{1},i_{2}a_{2},\ldots,
i_{m}c_{m}\ra_{R}:=b_{-c_{m},i_{m}}\ldots b_{-c_{2},i_{2}}b_{-c_{1},i_{1}}
{|R\ra}_{R}
\eea
with $b_{-a_{m},i_{m}}\in\Mns$  $a_{m+1}\geq a_{m}\geq 1/2$
 and $b_{-c_{m},i_{m}}\in\Mr$ $c_{m+1}\geq c_{m}\geq 0$
respectively where $m\in{\bf Z}_{\geq 0}$.

	 The observables are made out of bilinears in $b_{a,i}b_{c,i}$,\
restriction of the representations $\pns$ and $\pr$\ of $\sf Maj$ to
them produces reducible representations. To find the
inequivalent representations of the observable algebra $\bar {\cal A}$
note that $\sl Maj$ is the
\mbox{(antisymmetric)-product} of $M$ independent Majorana fields
(in contrast to \cite{fgv} where the Kac-Moody currents mix the colors
of the fermions), so to obtain the irreducible blocks we take the tensor
product of the corresponding decomposition of the Ising model, i.e.,
$\pns|_{\bar {\cal A}}={\pi}_{0}\oplus{\pi}_{1}$ see \cite{ms}:

\be
{({\pi}_{0}\oplus{\pi}_{1})}^{M}\leadsto
    \bigoplus_{r=0}^{M}\bigoplus_{(\ii)}
            {\pi}_{r(\ii)}
\ee
where $r(\ii)$ stands for $(i_{1},i_{2},\ldots,i_{r})$ with
$1\leq i_{1}<i_{2}<\cdots<i_{r}\leq M$ and $(0\leq r\leq M)$,
the second sum on the RHS for a fixed $r$ is over all possible
\mbox{r-tuples} $r(\ii)$ (there are $M!/r!(M-r)!$ \mbox{r-tuples});
the representations ${\pi}_{r(\ii)}$ have conformal weight $\fr{r}{2}$ and
are generated as modules by the following highest weights vectors of
the Virasoro algebra:

\bea
      &&|0\ra:=|NS\ra  \nn \\
&&|i-\med\ra_{NS}=b_{-\med,i}|0\ra\hspace{4cm}1\leq i\leq M \nn \\
&&|i_{2}-\med,i_{1}-\med\ra_{NS}=b_{-\med,i_{2}}b_{-\med,i_{1}}|0\ra
  \hspace{1.5cm}1\leq i_{1}<i_{2}\leq M \\
&&\quad\vdots \nn \\
&&|M\;-\med,M-1\;-\med,\ldots,1\;-\med\ra_{NS}=b_{-\med,M}b_{-\med,M-1}
\ldots b_{-\med,1}|0\ra  \nn \\
 \nn \\[1cm]
&&\mbox{where we used the notation in (\ref{vec1}) }    \nn \\
  \nn \\[1mm]
	&&L_{0}|0\ra=0 \nn \\
	&&L_{0}|i-\med\ra=\med|i-\med\ra \nn \\
	&&L_{0}|i_{1}-\med,i_{2}-\med\ra=|i_{1}-\med,i_{2}-\med\ra \\
	&&\vdots \nn \\
	&&L_{0}|M\;-\med,M-1\;-\med,\ldots,1\;-\med\ra=\fr{M}{2}
	                |M\;-\med,M-1\;-\med,\ldots,1\;-\med\ra   \nn
\eea

	And $\cal H_{NS}$ decomposes as follows

\be
 {\cal H}_{NS} = \bigoplus_{r=0}^{M}
   \bigoplus_{\ii}{\cal H}_{NS_{[r(\ii)]}} \label{nsrep}
\ee
Let ${\cal H}_{{NS}_{(0)}}\equiv {\cal H}_{0}$.
Turning now to the Rammond sector  ${\cal H}_{R}$ decomposes also
in $2^M$ independent vectors with conformal weight $\fr{M}{16}$, viz.:

\bea
                   |R\ra& \nn\\
	     b_{0,i}|R\ra& \; 1\le i \le M \nn\\
 b_{0,{i}_{2}}b_{0,{i}_{1}}|R\ra& \; 1\le i_{1}<i_{2}\le M \label{eqla} \\
	            \vdots& \\
b_{0,M}b_{0,M-1}\ldots b_{0,1}|R\ra& \nn
\eea

	Consequently ${\cal H}_{R}$ splits into:

\be  {\cal H}_{R}=\bigoplus_{i=1}^{2^M}{\cal H}_{R_{[r(\ii)]}}
              \label{rrep}
\ee
but now all representation spaces ${\cal H}_{R_{[r(\ii)]}}$ are equivalent.
Finally because ${\bar {\cal A}}\supset {\it Vir}$ the above
decompositions are a posteriori irreducible for $\bar {\cal A}$ (being
already irred.  for $\it Vir$). The representations generated by the
modules ${\cal H}_{NS_{[r(\ii)]}}$ and ${\cal H}_{R_{[r(\ii)]}}$
are ${\pi}_{r(\ii)}$
and by definition ${\pi}_{\fum}$ respectively.
 In particular the vacuum representation will be denoted by ${\pi}_{0}$.


\subsection{Morphisms}

	We define now auxiliary morphisms of the $\sf Maj$ algebra.
For $i,j \in \{1,2,\ldots M\}$

\be     {\rho}_{j}(b_{a,i})=b_{a,i}  \quad for\quad  j \neq i \ee
\be  {\rho}_{i}(b_{a,i})=\left\{ \begin{array}{ll}
                          ib_{a+\fum,i}  &a\geq\fum   \\
             \fr{i}{\sqrt{2}}(b_{\fum,i}-b_{-\fum,i}) & a=0 \\
                         -ib_{a-\fum,i}    & a\le -\fum
                                 \end{array}
           \right.  \label{endo} \ee

\bea
             \ai(b_{a,j}) &=&-b_{a,j} \quad for\quad j\neq i\nn\\
             \ai(b_{a,i}) &=&\left\{\begin{array}{ll}
                                  -b_{a,i} & a\neq 0, \pm\fum \\
                                   b_{-a,i} & a=0 , \pm\fum
                                  \end{array}
             \right. \label{auto}
\eea

	With these auxiliary constructs we build our endomorphism $\ro$:

\be  \ro = \prod_{i=1}^{M}\,{\rho}_{i}  \label{ro} \ee

\noindent
${\bf Remark}$\/{\em \quad
  The automorphirsms $\ai$ are inner automorphisms on $\Mns$
and on $\Mr$ respectively:
    \be {\alpha}_{j}(b_{a,i}) = U_{j}b_{a,i}U^{\ast}_{j} \nn \ee
with unitaries $U_{j}$
\be  U_{j}=(b_{\fum ,j} +  b_{-\fum ,j})\mbox{  for  }a\in {\bf Z}+\fum\quad
           b_{a,j}\in \Mns \label{U}	\ee
\be  U_{j}= \sqrt{2}b_{0,j}\mbox{  for  }a\in {\bf Z} \; b_{a,j}\in \Mr \ee
	On $\cal A$ the automorphirsms are not inner (\cite{ms}).}
\vspace{1cm}

  The endomorphism $\ro$ given by (\ref{ro}) is well-defined because the
${\rho}_{i}$`s commute among themselves and it is   obviously a
$\ast$-endomorphism. We will now consider arbitrary products of automorphisms
$\ai$ (there are a total of $2^{M}$) and denote an arbitrary
product as $\ag$.It is easy to calculate the action of $\ro$ on
the algebra of generators. We write here only the action on $L_{0}$ and
the central element $Y$:

\bea
      \ro(L_{0})&=&L_{0} - {\fum}N_{0} - \fr{M}{16}Y \nn \\
\smallskip \\
     {\ro}(Y)&=& -Y   \nn
\eea

	The irreducible representations of $\sf Maj$: $\pns$ and $\pr$
are intertwined by $\ro$ i.e., $\pns\circ\ro\cong\pr$; and being
the $\ag$  products of the automorphisms $\ai$ defined
in $(\ref{auto})$ they project also to inner automorphirsms of $\Mns$,
then it follows that $\pns\circ\ag\cong\pns$. Finally the vector $|NS\rangle$
is also a lowest weight of $\pns\circ\ro$ (see below).
	Now we want to understand how the morphisms $\ro$ and $\ag$
intertwine the representations of the observable algebra $\bar{\cal A}$.
First remember from (\ref{nsrep}) and from (\ref{rrep}) that the
Majorana algebra restricts to $\bar{\cal A}$ according to
$\pns\cong\bigoplus_{r=0}^{M}\bigoplus_{\ii}\pi_{r(\ii)}$ and
$\pr\cong\oplus 2^M\pi_{\fum}$, where the right-hand side means that $\pr$
decomposes into $2^{M}$ equivalent representations. Therefore by the above
discussion:

\be
\bigoplus_{r=0}^{M}\bigoplus_{\ii}\pi_{r(\ii)}\circ\ro\cong\pr\cong
\oplus 2^{M}\pi_{\fum} \label{comp}
\ee

	Consequently we have:

\be {\pi}_{r(\ii)}\circ\ro\cong\pi_{\fum} \ee

	In particular ${\pi}_{0}\circ\ro={\pi}_{\fum}$ and the conformal
weight of ${\pi}_{0}\circ\ro(L_{0})$ is:

\bean
  {\pi}_{0}\circ\ro(L_{0})|0\ra&=&\pi_{0}(L_{0}-\fum N_{0}-\fr{M}{16}
  Y)|0\ra=(\fr{M}{16})|0\ra \\
 \ro(L_{n})|0\ra&=&0 \quad{\rm for} \quad n>0
\eean

	The second equation above can be proven in the following way:
write the vacuum as $|0\ra\equiv |0\ra_{1}\times|0\ra_{2}\times\cdots
\times |0\ra_{M}$ where $|0\ra_{i}$ is the vacuum for ${\sf Maj}_{i}$
then for every $i$ we can apply the method of \cite{ms} to prove the
result.  	To determine what sector gives ${\pi}_{0}\circ\ag$
where $\ag$ is an arbitrary product of $\ai$ say
$\ag={\alpha}_{i_{1}}\circ
{\alpha}_{i_{2}}\circ\cdots\circ{\alpha}_{i_{r}}$ (with $1\leq i_{1}<
i_{2}<\cdots < i_{r}\leq M$) we calculate first the lowest weight of
${\pi}_{0}\circ\ag(L_{0})$ ,i.e. :

\[  {\pi}_{0}\circ\ag(L_{0})|0\ra=(L_{0}+\fr{n}{2}-{\sum}_{j}H_{\fum,j}
   -\fr{r}{4}[1+Y])|0\ra=\fr{n}{2}|0\ra \]

\noindent
where $H_{\fum,j}=b_{-\fum,j}b_{\fum,j}
       =J_{-\fum,\fum,j}+\fr{1}{4}[1-Y]$. Now
$\langle 0|{\pi}_{0}\circ\ag(A)|0\ra=\langle 0|U_{g}AU_{g}^{\ast}|0\ra
\linebreak
=\langle 0|{\pi}_{r(\ii)}(A)|0\ra$ ($U_{g}=U_{i_{1}}U_{i_{2}}\cdots
U_{i_{r}}$ with $U_{i}$ defined in (\ref{U})). The above computation
 gives us the result:
\[ {\pi}_{0}\circ\ag\equiv{\pi}_{0}\circ{\alpha}_{i_{1}}\circ
{\alpha}_{i_{2}}\circ\cdots\circ{\alpha}_{i_{r}}\cong{\pi}_{r(\ii)}\]


\section{Fusion Rules and Intertwining Operators}

	We prove now that the endomorphism $\ro$ defined in (\ref{ro})
satisfies the required fusion rules (\ref{fu1},\ref{fu2}). The equivalence
class of the representation ${\pi}_{0}\circ\rho$ of $\bar {\cal A}$
will be denoted $[\rho]$. Finally it will be verified that
$[\ro\ag]=[\ag\ro]$ the equivalence being given by the statistics operator
$\ep(\ro,\ag)$. In this and the following section we will always work in
the vacuum representation, therefore it will not be ambiguous to
 write $A$ for ${\pi}_{0}(A)$.
	Now for each  ${\rho}_{i}$ defined in (\ref{endo})
we can use the results of \cite{ms}:

\bea
	AR_{i}^{\ast}&=&R_{i}^{\ast}{\rho}_{i}^{2}(A) \nn \\
\smallskip \label{int}\\
	\ai(A)S_{i}^{\ast}&=&S_{i}^{\ast}{\rho}_{i}^{2} \nn
\eea

	Where $S_{i}^{\ast}=\ai(R_{i}^{\ast})$ and

\bean
 R_{i}^{\ast}&=&\ldots A_{3,i}A_{2,i}A_{1,i}P_{i}=(\prod_{n\geq1})P_{i}\\
 A_{n,i}&=&b_{n+\fum,i}b_{-n-\fum,i}+b_{-n+\fum,i}b_{n+\fum,i}
\eean
where $P_{i}=b_{\fum,i}b_{-\fum,i}$ is a projection, also
$R_{i}R_{i}^{\ast}=P_{i}$ and $R_{i}^{\ast}R_{i}=\bf 1$ ($\bf 1$ is the
identity operator).

	To decompose ${\rho}^{2}_{1/2}$ , we find the minimal
projectors in the commutant of ${\rho}^{2}_{1/2}$ in ${\cal
B(H}_{0})$. For this we calculate first its
action on the fermions:

\be
	{\rho}^{2}_{1/2}(b_{a,i})=\left\{ \ba{ll}
	                      -b_{a+1,i}, & a>0 \\
	  \smallskip &\hspace{2cm}\mbox{for $a\in{\bf Z}+\fum$}\\
	                      -b_{a-1,i}, & a<0
	\ea     \right.  \ee
\noindent
and ${\rho}^{2}_{1/2}(b_{a,i})=0$ $\forall i$ in the vacuum representation
${\pi}_{0}$, for $n\in {\bf Z}$. Then it is not difficult to see that the
intertwiner between ${\rho}^{2}_{1/2}$ and the vacuum,
call it $R$, must be the
product of all $R_{i}$ defined (\ref{int}):

\bea
	R&=&\prod_{i=1}^{M}R_{i} \nn \\
	\smallskip \label{R} \\
	AR^{\ast}&=&R^{\ast}{\rho}^{2}_{1/2}(A) \nn
\eea

	 For $A=b_{a,i}b_{c,i}$ (\ref{R}) can be verified by explicit
computation and consequently extends to $A\in\bar {\cal A}$. $R$ is
also a partial isometry:

\be P_{0}:=RR^{\ast}=\prod_{i=1}^{M}P_{i} \label{P} \ee

	Now apply all possible products of automorphisms $\ai$ from
(\ref{auto}) with different colors , there are a total of $2^{M}$
products, to $R^{\ast}$ obtaining $2^{M}$ intertwining operators,
call a generic element of this set $S_{g}^{\ast}$, in symbols

\bea
	\prod_{i=1}^{M}(1+\ai)&=&\sum_{g}\ag \nn \\
	\ag(R^{\ast})&=&S_{g}^{\ast} \label{S}
\eea

They satisfy

\[   \ag(A)S_{g}^{\ast}=S_{g}^{\ast}{\rho}^{2}_{1/2}(A)  \]

Because $\ag\circ\ro=\ro$ (see below).
\\[5mm]
\noindent
${\bf Proposition\/:}${\em \quad
The endomorphirsm $\ro$ satisfies the fusion rules

\be	[\ro\circ\ro]=\sum_{g}[\ag]  \ee
\noindent
where the sum goes over $2^{M}$ terms and the sub-index $g$ is an
abbreviation for a muti-index $\{i_{1}i_{2}\ldots ,i_{r}\}$.
where $r$ goes from $1$ to $M$ and the $i_{n}$ are one~by~one different
by convention $r=0$ represents the identity representation $id$ .
The automorphirsms $\ag$ are self-conjugate:

\[ {[\ag]}^{2}=[id.] \]

Finally we have

\be [\ag\circ\ro]=[\ro\circ\ag]=[\ro]  \ee}

{\em Proof.} It was proven above that $R$ intertwines between
${\rho}^{2}_{1/2}$ and the vacuum, suppose that $\ag\circ\ro=\ro$ (we
will prove it in a moment), then the $S_{g}$ defined in (\ref{S})
intertwine between ${\rho}^{2}_{1/2}$ and $\ag$ so we only have to
prove that we have a complete decomposition of ${\rho}^{2}_{1/2}$ this
is proven if the minimal projectors in the commutant of
${\rho}^{2}_{1/2}$ in ${\cal B(H}_{0})$ span the identity operator in
${\cal H}_{0}$ , by the definition of the $\ag$'s:

\[ \sum_{g}\ag(P_{0})=\prod_{i=1}^{M}(1+\ai )\cdot(P_{0})=
   \prod_{i=1}^{M}(1+\ai )\cdot(\prod_{i=1}^{M}P_{i})=1 \]

It is also clear that ${\ag}^{2}=1$ because of the self-~conjugacy
of the $\ai$'s. We prove now the last assertion of the prop. By
computation

\[ \ag\circ\ro(b_{a,i})=(-)^{g}\ro(b_{a,i}) \mbox{   for  $a\in{\bf Z}$} \]

\noindent
and in the vacuum representation both sides vanish for $a\in{\bf Z}+\fum$.
The $(-)^{g}$ is the grade of $\ag$ (the number of $\ai$'s in $\ag$). We
have also $\ag\circ\ro(Y_{i})=\ro(Y_{i})=-Y_{i}$. Therefore
$\ag \circ\ro=\ro$ on bilinears thus on $\cal A$ too. Now

\[ \ro\circ\ag(b_{a,i})=\left\{ \ba{ll}
                  (-)^{g}ib_{a+\fum,i}\quad    &a>0 \\
              (-)^{g}\fr{i}{\sqrt{2}}(b_{\fum,i}-b_{-\fum,i})\quad   &a=0\\
	           (-)^{g}ib_{a-\fum,i}\quad     &a<0
	    \ea \right. \]

\noindent
$\ro\circ\ag(b_{a,i})=0$ on the vacuum representation for $a\in{\bf Z}+\fum$,
while $\ro\circ\ag(Y_{i})=-Y_{i}=\ro(Y_{i})$. Let $V_{g}^{\ast}$ :
${\cal H}_{0}\rightarrow{\cal H}_{0}$ be a unitary operator such that

\be \ag\circ\ro(A)V_{g}^{\ast}=V_{g}^{\ast}\ro\ag(A) \ee

\noindent
for $A\in {\cal A}$, $V_{g}^{\ast}=\prod_{i\in g}(1-2b_{-\fum,i}b_{\fum,i})$
where the product is over the decomposition of $\ag$ in basis automorphirms
$\ai$. Note that $V_{g}^{\ast}\equiv {\ep}(\ro,\ag)$.

\section{The Statistical Operator}

  In this section $\rho =\ro$. It is easy to generalize the
expressions for the Jones projectors given in \cite{ms} to
our model, in the vacuum representation we have:

\bea &E_{2n+1}=\prod_{i=1}^{M}b_{\fum+n,i}b_{-\fum-n,i}\,,\nn\\
  \smallskip
   &E_{2n+2}=2^{-M}\prod_{i=1}^{M}[1-(b_{\fum+n,i}-b_{-\fum-n,i})
                          (b_{\fr{3}{2}+n,i}+b_{-\fr{3}{2}-i})]\nn
\eea
	This projections satisfy the Jones relations:
\bea
      	&E_{n}E_{m}=E_{m}E_{n},\qquad |n-m|\geq 2,\nn \\
        \smallskip \label{jo}
	&E_{n}E_{n\pm 1}E_{n}=d_{\rho}^{-2}E_{n} \nn
\eea
$d_{\rho}=2^{\fr{M}{2}}$ is the statistical dimension of $\rho$ and
$E_{1}\equiv P_{0}$ (see (\ref{P})). Let now $E_{g} =\ag(E_{1})$
(with the same conventions for $\ag$ as before), the statistical operator
${\ep}_{\rho}\equiv\ep(\rho,\rho)$ is a linear combination of this
projections note that $\eg=S_{g}S_{g}^{\ast}$
($S_{g}$ defined in (\ref{S})):

\be {\ep}_{\rho}=C\sum_{\ag}q(\ag)\eg
\ee
where the sum has $2^{M}$ terms and $q(id)=1$, $g=0$ representing the
identity, $q(\ag)^{2}=e^{2\pi ih_{\ag}}$ and the
constant $C^{2}=e^{-4\pi ih_{\rho}}$ (see \cite{fk}).
The $q\; : \;stab(\rho)\rightarrow {\bf Z}_{4}$ satisfy:

\be q(\ag)q({\alpha}_{f})={\sigma}(\ag,{\alpha}_{f})
    q(\ag\cdot{\alpha}_{f})
\ee
In our case ${\sigma}(\ag,\ag)=(-1)^{(g)}$ by the spin-statistics
theorem, see for example \cite{frs} and
${\sigma}(\ag,{\alpha}_{f})=\pm 1$, it is equal to $+1$ when $\ag$
and ${\alpha}_{f}$ have in their decomposition no terms or
an even number of $\ai$`s in common , otherwise it is equal to $-1$,
as it must be $\sigma$ as defined is a bihomomorphism.The values of
$C$ and $q(\ag)$ are obtained by applying the left-inverse $\Phi$ of
$\rho$ to the statistical operator: ${\Phi}(\ep(\rho,\rho))=
e^{2\pi ih_{\rho}}/d_{\rho}$, $h_{\rho}$ is the conformal dimension
of the sector $[\rho]$ i.e. $M/16$,
\noindent
this was done in \cite{fk}, in our
notation:

\[ e^{2\pi ih_{\rho}}C^{-1}=\fr{1}{d_{\rho}}\sum_{\ag}q(\ag)  \]
\noindent
with $C=e^{-\pi iM/8}$ ; $q(\ag)=e^{\pi i(g)/2}$ we can satisfy the
eq. above ($\rho$ is a real sector) and this are the only possible
solutions because
for $M=1$ we have the values found in \cite{ms}.\medskip

{\bf Acknowledgments}
	I thank Prof. B.Schroer and C.P.Staszkiewicz for discussions.



\begin{thebibliography}{9}

\bibitem{ms}Mack G.,Schomerus V.: Conformal Field Algebras with Quantum
Symmetry from the Theory of Superselectionsectors, Commun. Math. Phys.
${\bf 134}$, 139-196 (1990)

\bibitem{fk}Fr\"ohlich J. ,Kerler T.: Quantum Groups Quantum
Categories and Quantum Field Theory, [Springer Lectures Notes in
Mathematics](Springer Verlag, Berlin, in press), pg. 216

\bibitem{fgv}Fuchs J.,Ganchev A.,Vercserny\'es P.: Level 1 WZW Superselection
Sectors, Commun. Math. Phys.${\bf 146}$, 553-583  (1992)

\bibitem{frs1}FredenhagenK.,Rehren K.-H.,Schroer B.: Superselection sectors
with braid group statistics and exchange algebras. I: General Theory,
Commun. Math. Phys. ${\bf 125}$, 201-226 (1989)

\bibitem{frs}Fredenhagen K.,Rehren K.-H.,Schroer B.: Superselection Sectors
with Braid Group Statistics and Exchange Algebras II: Geometric Aspects and
Conformal Covariance, Rev.Math.Phys. [special issue] 113-157 (1992)

\end{thebibliography}
\end{document}